\documentclass[11pt]{article}

\usepackage{PhysRep}
\usepackage{Lep2Rep}
\usepackage{gg}
\usepackage{ff}
\usepackage{4f_s06}
\usepackage{smat}
\usepackage{gc}
\usepackage{mw}

\usepackage[english]{babel}
\usepackage{graphicx,rotating}
\usepackage{a4p,here}
\usepackage{cite,mcite,epsfig}

\renewcommand{\Huge}{\huge}
\newcommand{\updates}[1]%
{\fbox{\parbox{\linewidth}{\textbf{Updates with respect to summer 2006:}\\#1}}}

\input{rotate}

\begin{document}
\flushbottom
\begin{titlepage}
\begin{center}
\Large {EUROPEAN ORGANIZATION FOR NUCLEAR RESEARCH}
\end{center}

\begin{flushright}
       CERN-PH-EP/2007-039 \\
       LEPEWWG/2007-01  \\
       ALEPH 2007-001 PHYSICS 2007-001 \\
       DELPHI 2007-002 PHYS 949 \\
       L3 Note 2834   \\
       OPAL PR 426    \\
       arXiv:0712.0929 [hep-ex] \\
       {6 December 2007}
\end{flushright}

\begin{center}
\boldmath
\Huge {\bf %
           Precision 
           Electroweak Measurements and \\
           Constraints on the Standard Model\\[.5cm]

}
\unboldmath

\vspace*{1.0cm}
\Large {\bf
The LEP Collaborations
ALEPH, DELPHI, L3, OPAL, and \\ the LEP Electroweak Working
Group\footnote{ WWW access at {\tt http://www.cern.ch/LEPEWWG}

The members of the 
LEP Electroweak Working Group 
who contributed significantly to this
note are: \\
J.~Alcaraz,         %
A.~Bajo-Vaquero,    %
E.~Barberio,        %
D.~Bourilkov,       %
P.~Checchia,        %
R.~Chierici,        %
R.~Clare,           %
J.~D'Hondt,         %
B.~de~la~Cruz,      %
P.~de~Jong,         %
G.~Della~Ricca,     %
M.~Dierckxsens,     %
D.~Duchesneau,      %
G.~Duckeck,         %
M.~Elsing,          %
M.W.~Gr\"unewald,   %
A.~Gurtu,           %
J.B.~Hansen,        %
R.~Hawkings,        %
St.~Jezequel,       %
R.W.L.~Jones,       %
T.~Kawamoto,        %
E.~Lan{\c c}on,     %
W.~Liebig,          %
L.~Malgeri,         %
S.~Mele,            %
M.N.~Minard,        %
K.~M\"onig,         %
C.~Parkes,          %
U.~Parzefall,       %
B.~Pietrzyk,        %
G.~Quast,           %
P.~Renton,          %
S.~Riemann,         %
K.~Sachs,           %
A.~Straessner,      %
D.~Strom,           %
R.~Tenchini,        %
F.~Teubert,         %
M.A.~Thomson,       %
S.~Todorova-Nova,   %
A.~Valassi,         %
A.~Venturi,         %
H.~Voss,            %
C.P.~Ward,          %
N.K.~Watson,        %
P.S.~Wells,         %
St.~Wynhoff $^\dagger$. \\ %
$^\dagger$ deceased.
}\\
}
\vskip 0.5cm
\large\textbf{Prepared from Contributions %
to the 2007 Summer Conferences.}\\
\end{center}
\vfill
\begin{abstract}

This note presents constraints on Standard Model parameters using
published and preliminary precision electroweak results measured at
the electron-positron colliders LEP and SLC.
  The results are compared with precise electroweak measurements from
  other experiments, notably CDF and D\O\ at the Tevatron.
  Constraints on the input parameters of the Standard Model are
  derived from the results obtained in high-$Q^2$ interactions, and
  used to predict results in low-$Q^2$ experiments, such as atomic
  parity violation, M{\o}ller scattering, and neutrino-nucleon
  scattering.

\end{abstract}
\vfill
\end{titlepage}
\setcounter{page}{2}
\renewcommand{\thefootnote}{\arabic{footnote}}
\setcounter{footnote}{0}

\boldmath
\unboldmath

\section{Introduction}

The experimental results used here consist of the final and published
Z-pole results~\cite{bib-Z-pole} measured by the ALEPH, DELPHI, L3,
OPAL and SLC experiments, taking data at the electron-positron
colliders LEP and SLC. In addition, published and preliminary results
on the mass of the W boson, measured at {\LEPII} and the Tevatron, and
the mass of the top quark, measured at the Tevatron only, are
included.\footnote{ Since our last report~\cite{bib-EWEP-06}, the
following results have been published which are not included in the
results used in the
following:~\cite{DELPHI-LEP2:NTGC,DELPHI-LEP2:Zg,DELPHI-LEP2:CR,DELPHI-LEP2:TAUPOL}.}

The measurements %
allow to check the validity of the Standard Model (SM) and, within its
framework, to infer valuable information about its fundamental
parameters. The accuracy of the W- and Z-boson measurements makes them
sensitive to the mass of the top quark $\Mt$, and to the mass of the
Higgs boson $\MH$ through loop corrections. While the leading $\Mt$
dependence is quadratic, the leading $\MH$ dependence is logarithmic.
Therefore, the inferred constraints on $\Mt$ are much stronger than
those on $\MH$.

\section{Measurements}

The measurement results considered here are reported in
Table~\ref{tab-SMIN}.  Also shown are their predictions based on the
results of the SM fit to these combined high-$Q^2$ measurements,
reported in the last column of Table~\ref{tab-BIGFIT}.  The
measurements obtained at the Z pole by the LEP and SLC experiments
ALEPH, DELPHI, L3, OPAL and SLD, and their combinations, reported in
parts a), b) and c) of Table~\ref{tab-SMIN}, are final and
published~\cite{bib-Z-pole}.

The results on the W-boson mass by UA2~\cite{UA2-MW},
CDF~\cite{CDF-MW-PRL90, *CDF-MW-PRD90, *CDF-MW-PRL95, *CDF-MW-PRD95,
*CDF-MW-2000} and D\O~\cite{D0-MW:central, *D0-MW:endcap, *D0-MW:edge,
*D0-MW:large} in Run-I, and the W-boson width by CDF\cite{CDF-GW} and
D\O\cite{D0-GW} in Run-I, are combined by the Tevatron Electroweak
Working Group based on a detailed treatment of common systematic
uncertainties~\cite{PP-MW-GW:combination}. Including also the recent
result on $\MW$ based on Run-II data by CDF~\cite{CDF2MW}, the
combined results are: $\MW = 80429\pm39~\MeV$, $\GW =
2078\pm87~\MeV$.\footnote{The combined Tevatron result quoted for
$\GW$ and used here does not take into account the very recently
published CDF result based on Run-II data, $\GW =
2.032\pm0.073~\GeV$~\cite{CDF2GW}.}  Combining these results with the
preliminary {\LEPII} combination~\cite{bib-EWEP-06},
the resulting averages used here are:
\begin{eqnarray}
\MW & = & 80.398 \pm 0.025~\GeV\\
\GW & = &  2.140 \pm 0.060~\GeV\,.
\end{eqnarray}

For the mass of the top quark, $\Mt$, the published Run-I results from
CDF~\cite{Mtop1-CDF-di-l-PRLa, *Mtop1-CDF-di-l-PRLb,
*Mtop1-CDF-di-l-PRLb-E, *Mtop1-CDF-l+j-PRL, *Mtop1-CDF-l+j-PRD,
*Mtop1-CDF-all-j-PRL} and D\O~\cite{D0-top:prl-ll, *D0-top:prd-ll,
*D0-top:prl-lj, *D0-top:prd-lj, *Mtop1-D0-l+j-new1,
*Mtop1-D0-l+j-new2, *Mtop1-D0-all-j-PRL}, and recent preliminary and
published results based on Run-II data from
CDF~\cite{Mtop2-CDF-di-l-1fbPRD, Mtop2-CDF-lxy-new, Mtop2-CDF-l+j-new,
Mtop2-CDF-all-j-new} and
D\O~\cite{Mtop2-D0-l+j-new,Mtop2-D0-di-l-new}, are combined by the
Tevatron Electroweak Working Group with the result:
$\Mt=170.9\pm1.8~\GeV$~\cite{TeVEWWGtop-0703}.

In addition, the following final results obtained in low-$Q^2$
interactions and reported in Table~\ref{tab-SMpred} are considered:
(i) the measurements of atomic parity violation in
caesium\cite{QWCs:exp:1, QWCs:exp:2}, with the numerical
result\cite{QWCs:theo:2003:new} taken from a recently published
revised analysis of QED radiative corrections applied to the raw
measurement; (ii) the result of the E-158 collaboration on the
electroweak mixing angle\footnote{ E-158 quotes in the MSbar scheme,
evolved to $Q^2=\MZ^2$.  We add 0.00029 to the quoted value in order
to obtain the effective electroweak mixing angle~\cite{PDG2004}.}
measured in M{\o}ller scattering~\cite{E158RunI, *E158RunI+II+III};
and (iii) the final result of the NuTeV collaboration on
neutrino-nucleon neutral to charged current cross section
ratios~\cite{bib-NuTeV-final}.

Using neutrino-nucleon data with an average $Q^2\simeq20~\GeV^2$, the
NuTeV collaboration has extracted the left- and right-handed couplings
combinations 
$\gnlq^2=4\gln^2(\glu^2+\gld^2) =
[1/2-\swsqeff+(5/9)\swsqsqeff]\rhon\rho_{\mathrm{ud}}$ and
$\gnrq^2=4\gln^2(\gru^2+\grd^2) =
(5/9)\swsqsqeff\rhon\rho_{\mathrm{ud}}$: 
$\gnlq^2=0.30005\pm0.00137$ and $\gnrq^2=0.03076\pm0.00110$, with a
correlation of $-0.017$.  While the result on $\gnrq$ agrees with the
$\SM$ expectation, the result on $\gnlq$, measured nearly eight times
more precisely, shows a deficit with respect to the expectation at the
level of 2.8 standard deviations.

An additional input parameter, not shown in the table, is the Fermi
constant $G_F$, determined from the $\mu$ lifetime, $G_F = 1.16637(1)
\cdot 10^{-5}~\GeV^{-2}$\cite{bib-Gmu-1, *bib-Gmu-2, *bib-Gmu-3}.
Recent new measurements of $G_F$ yield values which are in good
agreement~\cite{Chitwood:2007pa,Barczyk:2007hp}.  The relative error
of $G_F$ is comparable to that of $\MZ$; both errors have negligible
effects on the fit results.

\section{Theoretical and Parametric Uncertainties}

Detailed studies of the theoretical uncertainties in the SM
predictions due to missing higher-order electroweak corrections and
their interplay with QCD corrections had been carried out by the
working group on `Precision calculations for the $\Zzero$
resonance'\cite{bib-PCLI}, and later in~\cite{BP:98,PCP99}.
Theoretical uncertainties are evaluated by comparing different but,
within our present knowledge, equivalent treatments of aspects such as
resummation techniques, momentum transfer scales for vertex
corrections and factorisation schemes.  The effects of these
theoretical uncertainties are reduced by the inclusion of higher-order
corrections\cite{bib-twoloop,bib-QCDEW} in the electroweak libraries
TOPAZ0~\cite{Montagna:1993py, *Montagna:1993ai, *Montagna:1996ja,
*Montagna:1998kp} and ZFITTER~\cite{Bardin:1989di, *Bardin:1990tq,
*Bardin:1991fu, *Bardin:1991de, *Bardin:1992jc, *Bardin:1999yd,
*Kobel:2000aw, *Arbuzov:2005ma}.

The use of the higher-order QCD corrections\cite{bib-QCDEW} increases
the value of $\alfmz$ by 0.001, as expected.  The effects of missing
higher-order QCD corrections on $\alfmz$ covers missing higher-order
electroweak corrections and uncertainties in the interplay of
electroweak and QCD corrections. A discussion of theoretical
uncertainties in the determination of $\alfas$ can be found in
References~\citen{bib-PCLI} and~\citen{bib-SMALFAS}, with a recent
analysis in Reference~\citen{Stenzel:2005sg} where the theoretical
uncertainty is estimated to be about 0.001 for the analyses presented
in the following.

Recently, the complete (fermionic and bosonic) two-loop corrections
for the calculation of $\MW$~\cite{Twoloop-MW}, and the complete
fermionic two-loop corrections for the calculation of
$\swsqeffl$~\cite{Twoloop-sin2teff} have been calculated.  Including
three-loop top-quark contributions to the $\rho$ parameter in the
limit of large $\Mt$~\cite{Threeloop-rho}, efficient routines for
evaluating these corrections have been implemented since version 6.40
in the semi-analytical program ZFITTER.  The remaining theoretical
uncertainties are estimated to be $4~\MeV$ on $\MW$ and 0.000049 on
$\swsqeffl$.  The latter uncertainty dominates the theoretical
uncertainty in SM fits and the extraction of constraints on the mass
of the Higgs boson presented below. For a complete picture, the
complete two-loop calculation for the partial Z decay widths should be
calculated.

The determination of the size of remaining theoretical uncertainties
is under continuous study.  The theoretical errors discussed above are
not included in the results presented in Tables~\ref{tab-BIGFIT}
and~\ref{tab-SMpred}.  At present the impact of theoretical
uncertainties on the determination of $\SM$ parameters from the
precise electroweak measurements is small compared to the error due to
the uncertainty in the value of $\alpha(\MZ^2)$, which is included in
the results.

The uncertainty in $\alpha(\MZ^2)$ arises from the contribution of
light quarks to the photon vacuum polarisation
($\Delta\alpha_{\mathrm{had}}^{(5)}(\MZ^2)$):
\begin{equation}
\alpha(\MZ^2) = \frac{\alpha(0)}%
   {1 - \Delta\alpha_\ell(\MZ^2) -
   \Delta\alpha_{\mathrm{had}}^{(5)}(\MZ^2) -
   \Delta\alpha_{\mathrm{top}}(\MZ^2)} \,,
\end{equation}
where $\alpha(0)=1/137.036$.  The top contribution, $-0.00007(1)$,
depends on the mass of the top quark, and is therefore determined
inside the electroweak libraries TOPAZ0 and ZFITTER.  The leptonic
contribution is calculated to third order\cite{bib-alphalept} to be
$0.03150$, with negligible uncertainty.

For the hadronic contribution, we no longer use the value $0.02804 \pm
0.00065$\cite{bib-JEG2,bib-Burk}, but rather the new evaluation
$0.02758\pm0.00035$~\cite{bib-BP05} which takes into account published
results on electron-positron annihilations into hadrons at low
centre-of-mass energies by the BES collaboration~\cite{BES_01}, as
well as the revised published results from CMD-2~\cite{CMD_03} and new
results from KLOE~\cite{KLOE_04}.  The reduced uncertainty still
causes an error of 0.00013 on the $\SM$ prediction of $\swsqeffl$, and
errors of 0.2~\GeV{} and 0.1 on the fitted values of $\Mt$ and
$\log(\MH)$, included in the results presented below.  The effect on
the $\SM$ prediction for $\Gll$ is negligible.  The $\alfmz$ values
from the $\SM$ fits presented here are stable against a variation of
$\alpha(\MZ^2)$ in the interval quoted.

There are also several evaluations of
$\Delta\alpha^{(5)}_{\mathrm{had}}(\MZ^2)$~\cite{bib-Swartz,
bib-Zeppe, bib-Alemany, bib-Davier, bib-alphaKuhn, bib-jeger99,
bib-Erler, bib-ADMartin, bib-Troconiz-Yndurain, bib-Hagiwara:2003,
bib-Troconiz-Yndurain-2004} which are more theory-driven.  The most
recent of these (Reference \citen{bib-Troconiz-Yndurain-2004}) also
includes the new results from BES, yielding $0.02749\pm0.00012$.  To
show the effects of the uncertainty of $\alpha(\MZ^2)$, we also use
this evaluation of the hadronic vacuum polarisation.  Note that all
these evaluations obtain values for
$\Delta\alpha^{(5)}_{\mathrm{had}}(\MZ^2)$ consistently lower than -
but in agreement with - the old value of $0.02804 \pm 0.00065$.

\begin{table}[p]
\begin{center}
\renewcommand{\arraystretch}{1.10}
\begin{tabular}{|ll||r|r|r|r|}
\hline
 && \mcc{Measurement with}  &\mcc{Systematic} & \mcc{Standard} & \mcc{Pull} \\
 && \mcc{Total Error}       &\mcc{Error}      & \mcc{Model fit}&            \\
\hline
\hline
&&&&& \\[-3mm]
& $\Delta\alpha^{(5)}_{\mathrm{had}}(\MZ^2)$\cite{bib-BP05}
                & $0.02758 \pm 0.00035$ & 0.00034 &0.02768& $-0.3$ \\
&&&&& \\[-3mm]
\hline
a) & \underline{\LEPI}   &&&& \\
   & line-shape and      &&&& \\
   & lepton asymmetries: &&&& \\
&$\MZ$ [\GeV{}] & $91.1875\pm0.0021\pz$
                & ${}^{(a)}$0.0017$\pz$ &91.1875$\pz$ & $ 0.0$ \\
&$\GZ$ [\GeV{}] & $2.4952 \pm0.0023\pz$
                & ${}^{(a)}$0.0012$\pz$ & 2.4957$\pz$ & $-0.2$ \\
&$\shad$ [nb]   & $41.540 \pm0.037\pzz$ 
                & ${}^{(b)}$0.028$\pzz$ &41.477$\pzz$ & $ 1.7$ \\
&$\Rl$          & $20.767 \pm0.025\pzz$ 
                & ${}^{(b)}$0.007$\pzz$ &20.744$\pzz$ & $ 0.9$ \\
&$\Afbzl$       & $0.0171 \pm0.0010\pz$ 
                & ${}^{(b)}$0.0003\pz & 0.0165\pz     & $ 0.7$ \\
&+ correlation matrix~\cite{bib-Z-pole} &&&& \\
&                                             &&&& \\[-3mm]
&$\tau$ polarisation:                         &&&& \\
&$\cAl~(\ptau)$ & $0.1465\pm 0.0033\pz$ 
                & 0.0016$\pz$ & 0.1481$\pz$ & $-0.5$ \\
                      &                       &&&& \\[-3mm]
&$\qq$ charge asymmetry:                      &&&& \\
&$\swsqeffl(\Qfbhad)$
                & $0.2324\pm0.0012\pz$ 
                & 0.0010$\pz$ & 0.23138     & $ 0.8$ \\
&                                             &&&& \\[-3mm]
\hline
b) & \underline{SLD} &&&& \\
&$\cAl$ (SLD)   & $0.1513\pm 0.0021\pz$ 
                & 0.0010$\pz$ & 0.1481$\pz$ & $ 1.5$ \\
&&&&& \\[-3mm]
\hline
c) & \underline{{\LEPI}/SLD Heavy Flavour} &&&& \\
&$\Rbz{}$        & $0.21629\pm0.00066$  
                 & 0.00050     & 0.21586     & $ 0.7$ \\
&$\Rcz{}$        & $0.1721\pm0.0030\pz$
                 & 0.0019$\pz$ & 0.1722$\pz$ & $ 0.0$ \\
&$\Afbzb{}$      & $0.0992\pm0.0016\pz$
                 & 0.0007$\pz$ & 0.1038$\pz$ & $-2.9$ \\
&$\Afbzc{}$      & $0.0707\pm0.0035\pz$
                 & 0.0017$\pz$ & 0.0743$\pz$ & $-1.0$ \\
&$\cAb$          & $0.923\pm 0.020\pzz$
                 & 0.013$\pzz$ & 0.935$\pzz$ & $-0.6$ \\
&$\cAc$          & $0.670\pm 0.027\pzz$
                 & 0.015$\pzz$ & 0.668$\pzz$ & $ 0.1$ \\
&+ correlation matrix~\cite{bib-Z-pole} &&&& \\
&                                              &&&& \\[-3mm]
\hline
d) & \underline{{\LEPII} and Tevatron} &&&& \\
&$\MW$ [\GeV{}] ({\LEPII}, Tevatron)
& $80.398 \pm 0.025\pzz$ &      $\pzz$   & 80.374$\pzz$ & $ 1.0$ \\
&$\GW$ [\GeV{}] ({\LEPII}, Tevatron)
& $ 2.140 \pm 0.060\pzz$ &      $\pzz$   &  2.092$\pzz$ & $ 0.8$ \\
&$\Mt$ [\GeV{}] (Tevatron~\cite{TeVEWWGtop-0703})
& $170.9\pm 1.8\pzz\pzz$ & 1.5$\pzz\pzz$ & 171.3$\pzz\pzz$ & $-0.2$ \\
\hline
\end{tabular}\end{center}
\caption[]{ Summary of high-$Q^2$ measurements included in the
  combined analysis of SM parameters. Section~a) summarises {\LEPI}
  averages, Section~b) SLD results ($\cAl$ includes $\ALR$ and
  the polarised lepton asymmetries), Section~c) the {\LEPI} and SLD
  heavy flavour results, and Section~d) electroweak measurements from
  {\LEPII} and the Tevatron.  The total errors in column 2 include the
  systematic errors listed in column 3.  Although the systematic
  errors include both correlated and uncorrelated sources, the
  determination of the systematic part of each error is approximate.
  The $\SM$ results in column~4 and the pulls (difference between
  measurement and fit in units of the total measurement error) in
  column~5 are derived from the SM fit including all high-$Q^2$ data
  (Table~\ref{tab-BIGFIT}, column~4).\\ $^{(a)}$\small{The systematic
  errors on $\MZ$ and $\GZ$ contain the errors arising from the
  uncertainties in the $\LEPI$ beam energy only.}\\
  $^{(b)}$\small{Only common systematic errors are indicated.}\\ }
\label{tab-SMIN}
\end{table}

\section{Selected Results}

\begin{figure}[htbp]
\begin{center}
$ $ \vskip -1cm
  \mbox{\includegraphics[width=0.9\linewidth]{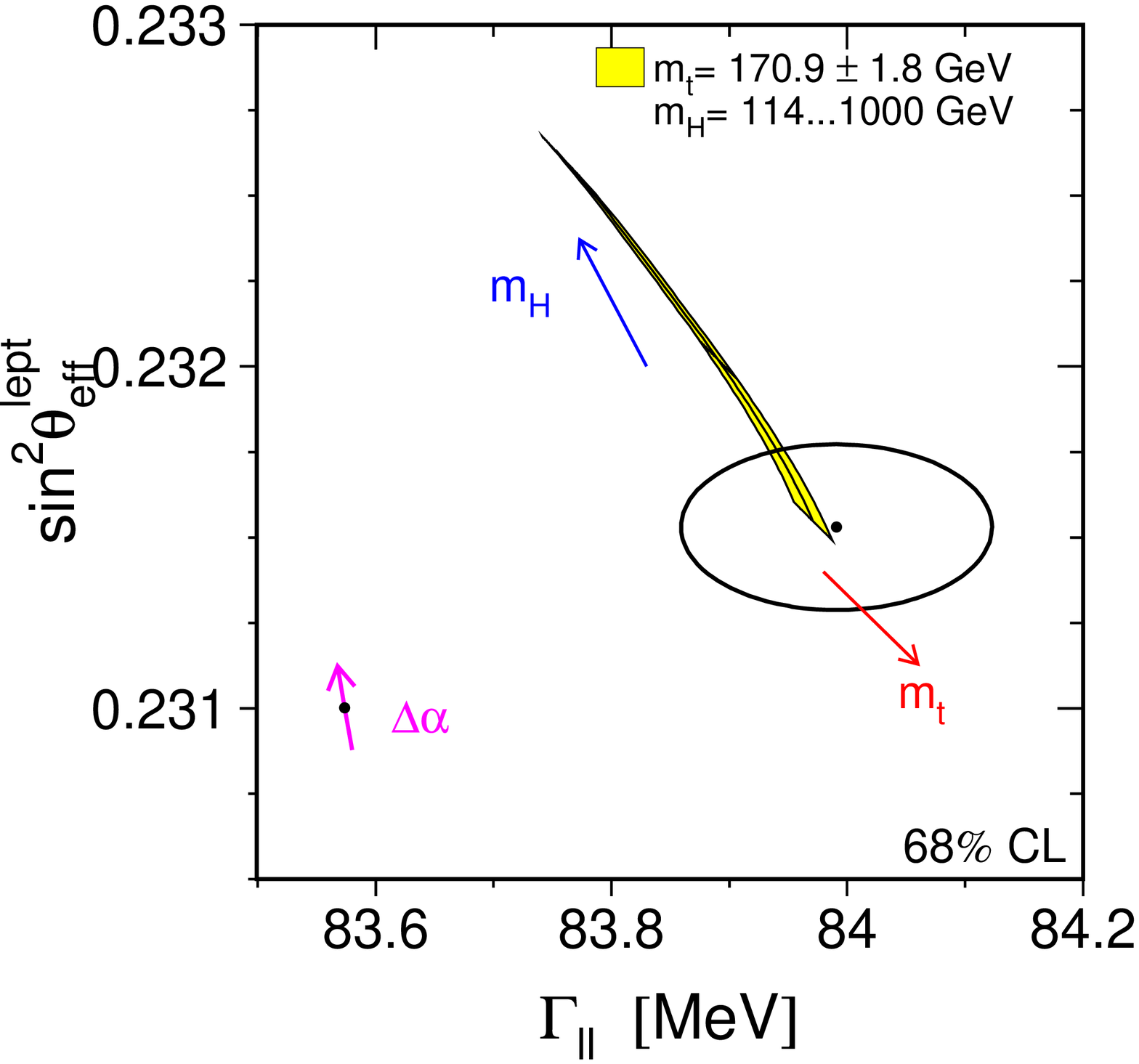}}
\end{center}
\vskip -1cm
\caption[]{ $\LEPI$+SLD measurements~\cite{bib-Z-pole} of $\swsqeffl$
  and $\Gll$ and the SM prediction.  The point shows the predictions
  if among the electroweak radiative corrections only the photon
  vacuum polarisation is included. The corresponding arrow shows
  variation of this prediction if $\alpha(\MZ^2)$ is changed by one
  standard deviation. This variation gives an additional uncertainty
  to the SM prediction shown in the figure.  }
\label{fig-gllsef}
\end{figure}

Figure~\ref{fig-gllsef} shows a comparison of the leptonic partial
width from {\LEPI}, $\Gll=83.985\pm0.086~\MeV$~\cite{bib-Z-pole}, and
the effective electroweak mixing angle from asymmetries measured at
{\LEPI} and SLD, $\swsqeffl=0.23153\pm0.00016$~\cite{bib-Z-pole}, with
the SM shown as a function of $\Mt$ and $\MH$.  Good agreement with
the $\SM$ prediction using the most recent measurements of $\Mt$ and
$\MW$ is observed.  The point with the arrow indicates the prediction
if among the electroweak radiative corrections only the photon vacuum
polarisation is included, which shows that the precision electroweak
Z-pole data are sensitive to non-trivial electroweak corrections.
Note that the error due to the uncertainty on $\alpha(\MZ^2)$ (shown
as the length of the arrow) is not much smaller than the experimental
error on $\swsqeffl$ from {\LEPI} and SLD.  This underlines the
continued importance of a precise measurement of
$\sigma(\mathrm{e^+e^-\rightarrow hadrons})$ at low centre-of-mass
energies.

Of the measurements given in Table~\ref{tab-SMIN}, $\Rl$ is one of the
most sensitive to QCD corrections.  For $\MZ=91.1875$~\GeV{}, and
imposing $\Mt=170.9\pm1.8$~\GeV{} as a constraint,
$\alfas=0.1221\pm0.0037$ is obtained.  Alternatively,
$\slept\equiv\shad/\Rl=2.0003\pm0.027~\nb$~\cite{bib-Z-pole} which has
higher sensitivity to QCD corrections and less dependence on $\MH$
yields: $\alfas=0.1177\pm0.0030$.  Typical errors arising from the
variation of $\MH$ between $100~\GeV$ and $200~\GeV$ are of the order
of $0.001$, somewhat smaller for $\slept$.  These results on $\alfas$,
as well as those reported in the next section, are in very good
agreement with world averages ($\alfmz=0.118 \pm
0.002$\cite{common_bib:pdg2000}, or $\alfmz=0.1178 \pm 0.0033$ based
solely on NNLO QCD results excluding the {\LEPI} lineshape results and
accounting for correlated errors~\cite{QCD:Bethke:2000,
*Bethke:2004uy}).

\clearpage

\section{Standard Model Analyses}

In the following, several different SM analyses as reported in
Table~\ref{tab-BIGFIT} are discussed.  The $\chi^2$ minimisation is
performed with the program MINUIT~\cite{MINUIT}, and the predictions
are calculated with ZFITTER as a function of the five SM input
parameters $\dalhad$, $\alfmz$, $\MZ$, $\Mt$ and $\LOGMH$ which are
varied simultaneously in the fits; see~\cite{bib-Z-pole} for details
on the fit procedure.  The somewhat large $\chi^2$/d.o.f.{} for all of
these fits is caused by the large dispersion in the values of the
leptonic effective electroweak mixing angle measured through the
various asymmetries at {\LEPI} and SLD~\cite{bib-Z-pole}.
Following~\cite{bib-Z-pole} for the analyses presented here, this
dispersion is interpreted as a fluctuation in one or more of the input
measurements, and thus we neither modify nor exclude any of them.  A
further significant increase in $\chi^2$/d.o.f.{} is observed when the
NuTeV results are included in the analysis.

\begin{figure}[htbp]
\begin{center}
\includegraphics[width=0.9\linewidth]{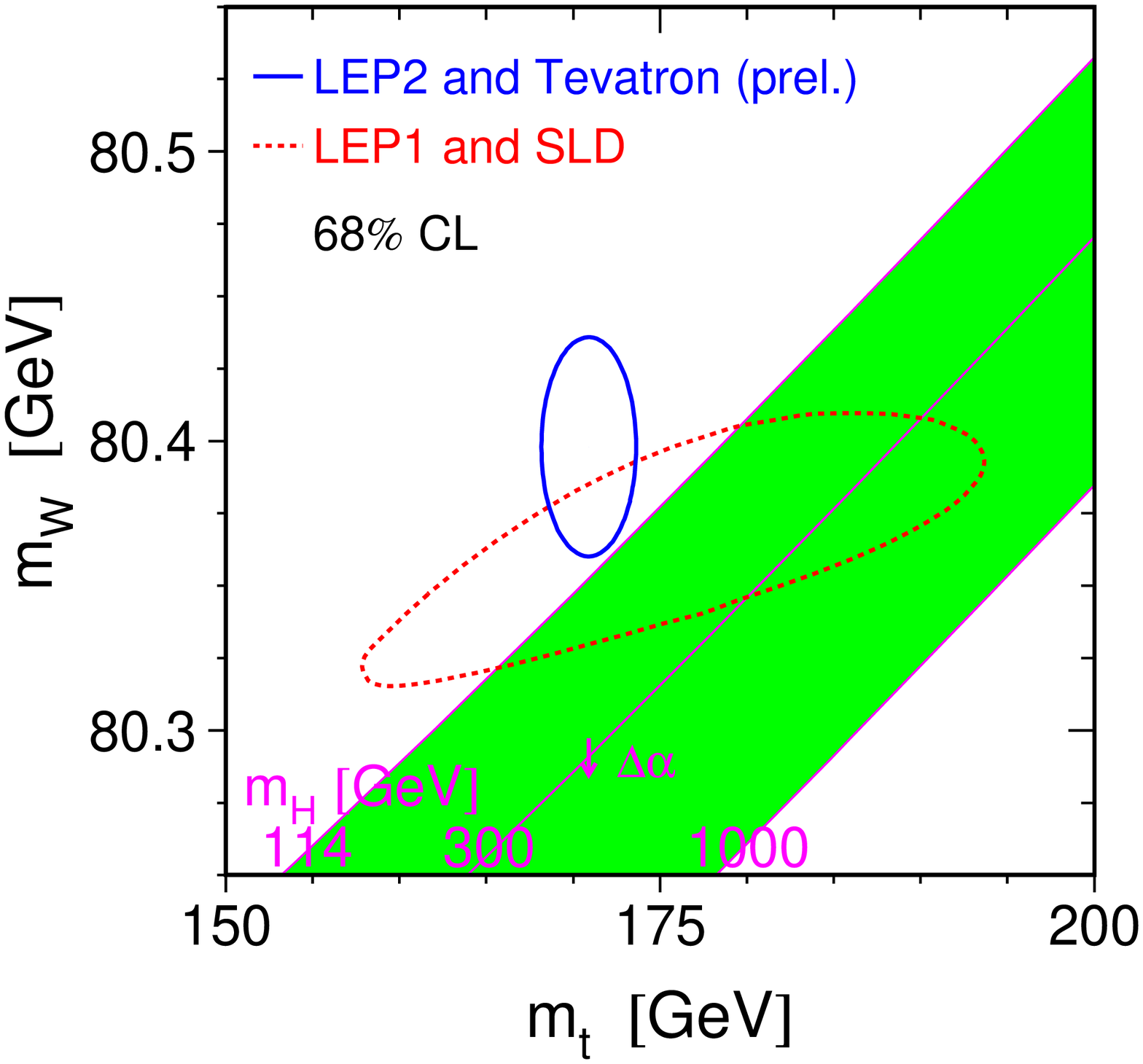}
\caption[]{ The comparison of the indirect measurements of $\MW$ and
  $\Mt$ ($\LEPI$+ SLD data) (solid contour) and the direct
  measurements ($\pp$ colliders and $\LEPII$ data) (dashed contour).
  In both cases the 68\% CL contours are plotted.  Also shown is the
  SM relationship for the masses as a function of the Higgs mass. The
  arrow labelled $\Delta\alpha$ shows the variation of this relation
  if $\alpha(\MZ^2)$ is changed by one standard deviation. This
  variation gives an additional uncertainty to the SM band shown in
  the figure.}
\label{fig:mtmW}
\end{center}
\end{figure}

To test the agreement between the Z-pole data~\cite{bib-Z-pole}
({\LEPI} and SLD) and the SM, a fit to these data is performed.  The
result is shown in Table~\ref{tab-BIGFIT}, column~1. The indirect
constraints on $\MW$ and $\Mt$ from this data sample are shown in
Figure~\ref{fig:mtmW}, compared with the direct measurements.  Also
shown are the SM predictions for Higgs masses between 114 and
1000~\GeV.  As can be seen in the figure, the indirect and direct
measurements of $\MW$ and $\Mt$ are in good agreement, and both sets
prefer a low value of the Higgs mass.

\begin{figure}[htbp]
\begin{center}
\includegraphics[width=0.9\linewidth]{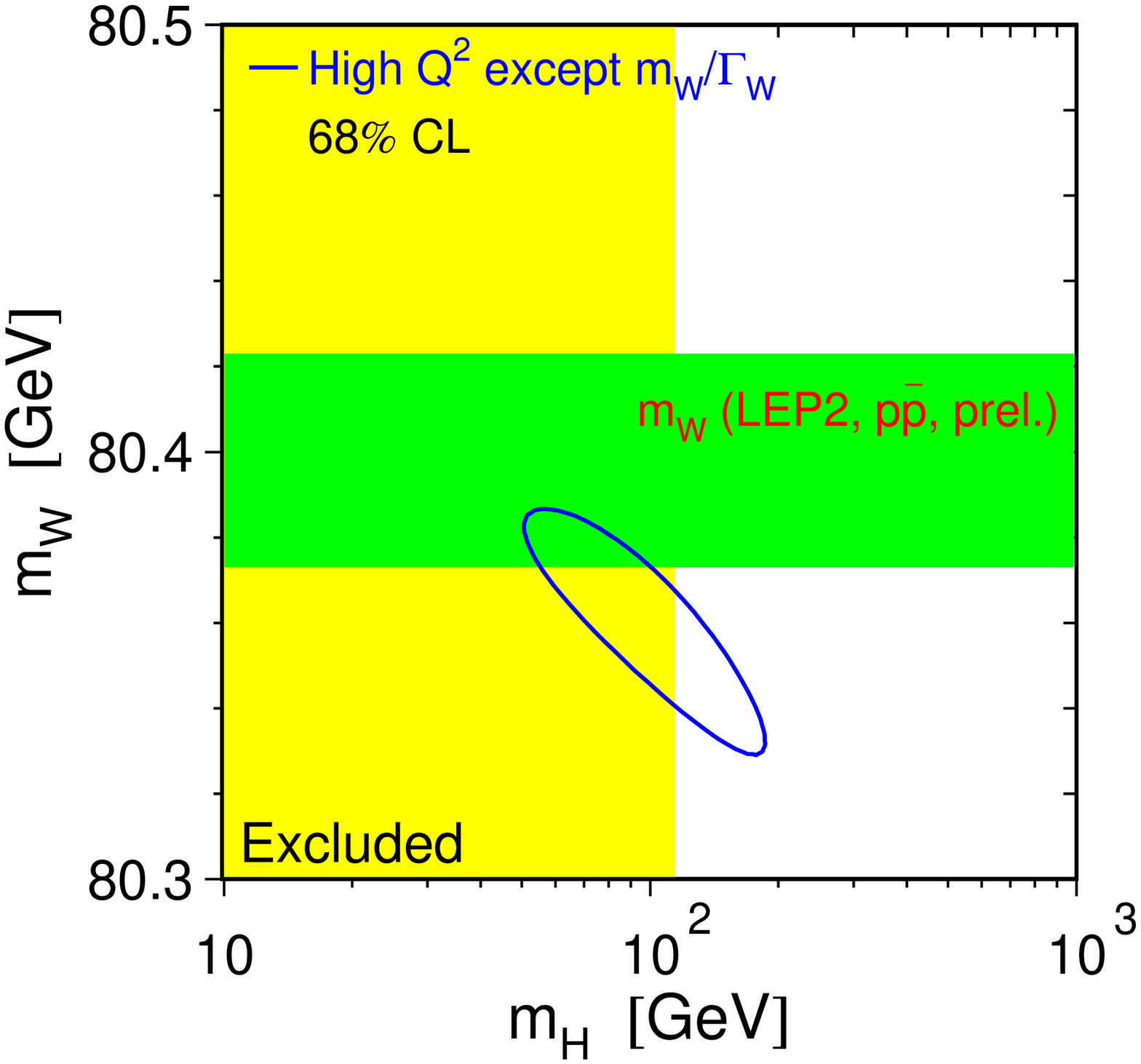}
\end{center}
\vspace*{-0.6cm}
\caption[]{
  The 68\% confidence level contour in $\MW$ and $\MH$ for the fit to
  all data except the direct measurement of $\MW$, indicated by the
  shaded horizontal band of $\pm1$ sigma width.  The vertical band
  shows the 95\% CL exclusion limit on $\MH$ from the direct search.
  }
\label{fig-mhmw}
\end{figure}

\begin{figure}[htbp]
\begin{center}
\includegraphics[width=0.9\linewidth]{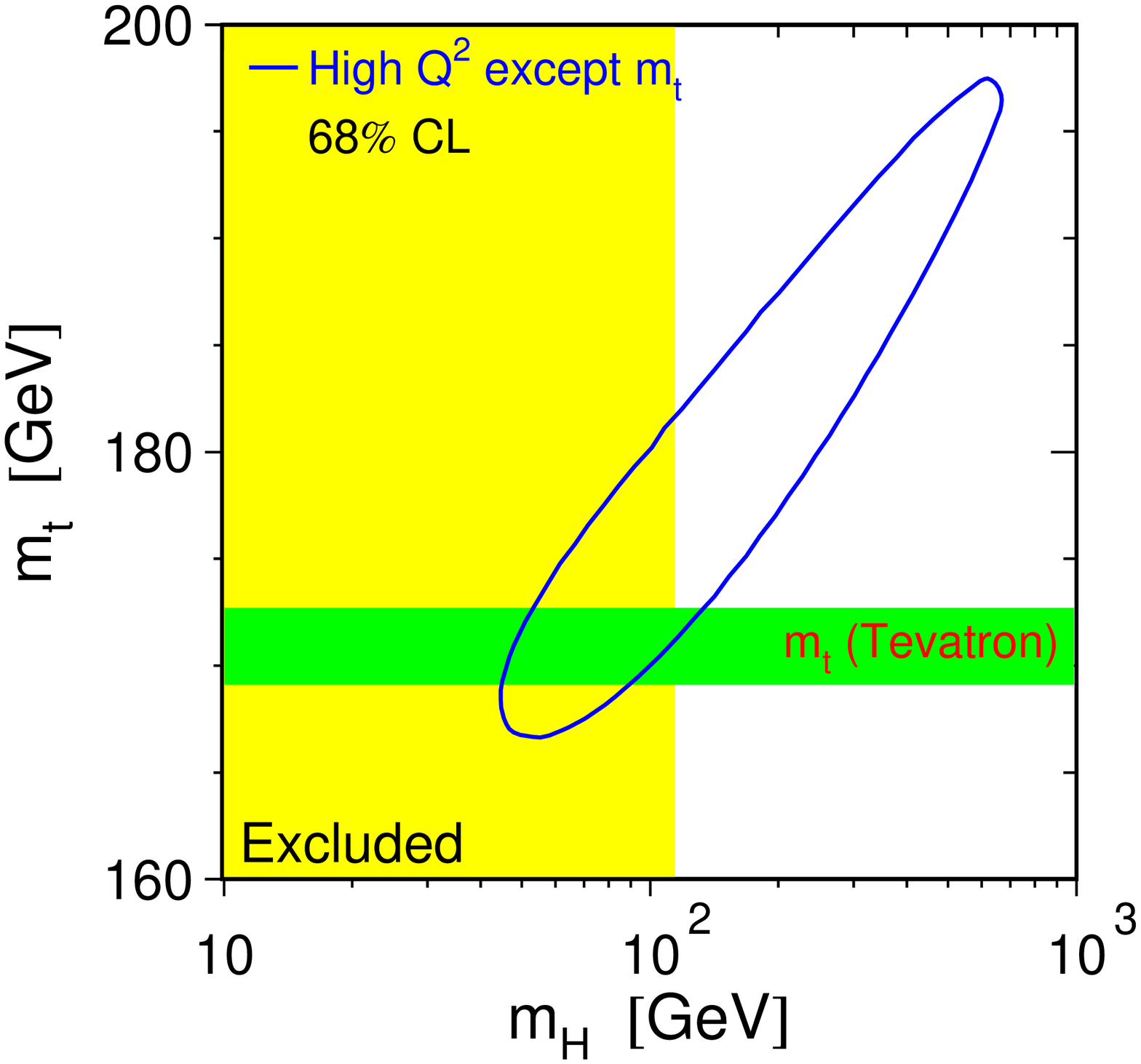}
\end{center}
\vspace*{-0.6cm}
\caption[]{ The 68\% confidence level contour in $\Mt$ and $\MH$ for
  the fit to all data except the direct measurement of $\Mt$,
  indicated by the shaded horizontal band of $\pm1$ sigma width.  The
  vertical band shows the 95\% CL exclusion limit on $\MH$ from the
  direct search.  }
\label{fig-mhmt}
\end{figure}

For the fit shown in column~2 of Table~\ref{tab-BIGFIT}, the direct
$\Mt$ measurement is included to obtain the best indirect
determination of $\MW$.  The result is also shown in
Figure~\ref{fig-mhmw}.  Also in this case, the indirect determination
of W boson mass, $80.360\pm0.020~\GeV$, is in good agreement with the
direct measurements from {\LEPII} and the Tevatron, $\MW=
80.398\pm0.025~\GeV$.  For the fit shown in column~3 of
Table~\ref{tab-BIGFIT} and Figure~\ref{fig-mhmt}, the direct $\MW$ and
$\GW$ measurements from {\LEPII} and the Tevatron are included instead
of the direct $\Mt$ measurement in order to obtain the constraint
$\Mt= 179^{+12}_{-9}~\GeV$, in good agreement with the direct
measurement of $\Mt = 170.9\pm1.8~\GeV$.

\begin{figure}[htbp]
\begin{center}
\includegraphics[width=0.9\linewidth]{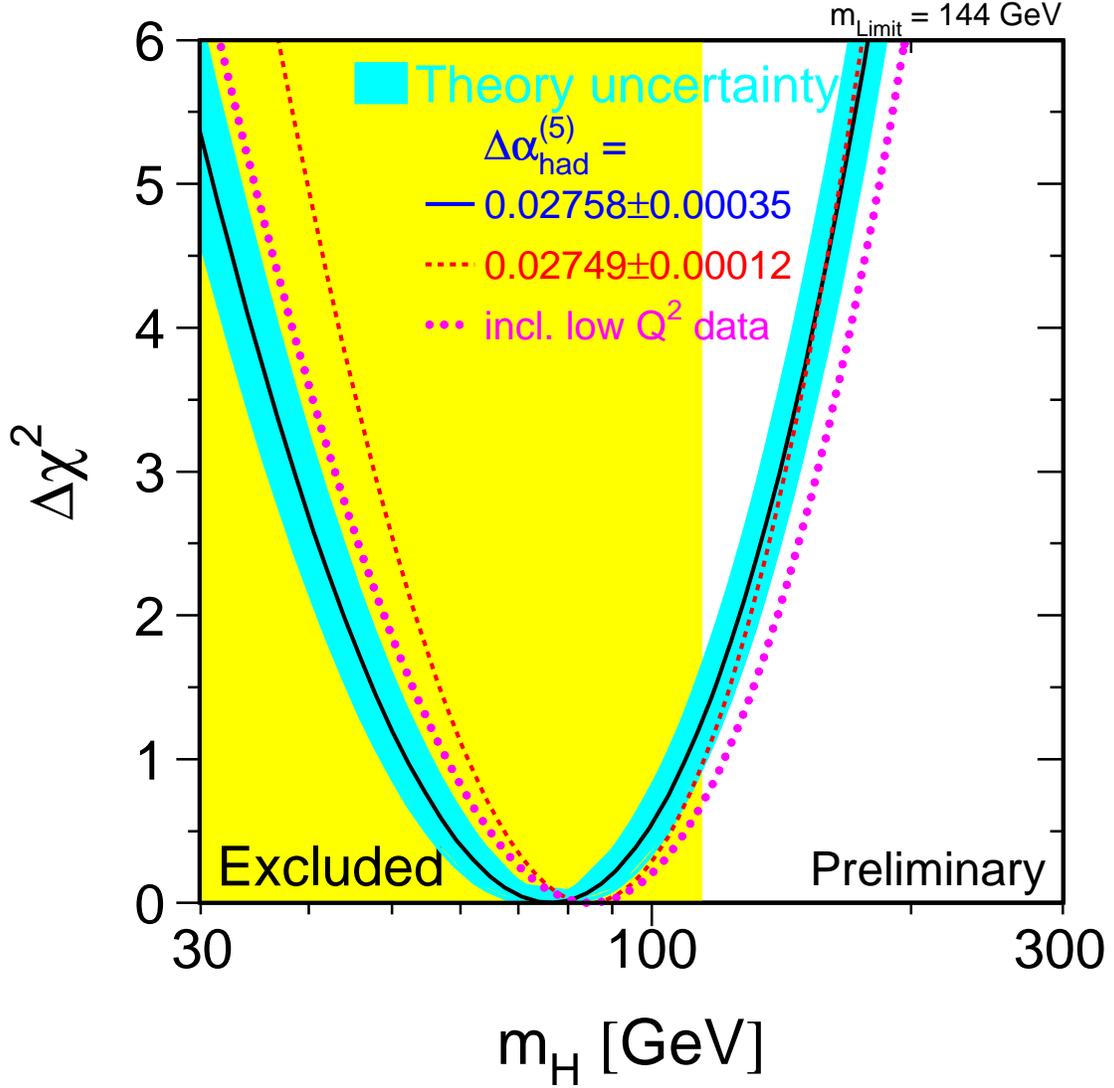}
\end{center}
\vspace*{-0.6cm}
\caption[]{ $\Delta\chi^{2}=\chi^2-\chi^2_{min}$ {\it vs.} $\MH$
  curve.  The line is the result of the fit using all high-$Q^2$ data
  (last column of Table~\protect\ref{tab-BIGFIT}); the band represents
  an estimate of the theoretical error due to missing higher order
  corrections.  The vertical band shows the 95\% CL exclusion limit on
  $\MH$ from the direct search.  The dashed curve is the result
  obtained using the evaluation of
  $\Delta\alpha^{(5)}_{\mathrm{had}}(\MZ^2)$ from
  Reference~\citen{bib-Troconiz-Yndurain-2004}. The dotted curve is
  the result obatined including also the low-$Q^2$ data. }
\label{fig-chiex}
\end{figure}

Finally, the best constraints on $\MH$ are obtained when all
high-$Q^2$ measurements are used in the fit.  The results of this fit
are shown in column~4 of Table~\ref{tab-BIGFIT}.  The predictions of
this fit for observables measured in high-$Q^2$ and low-$Q^2$
reactions are listed in Tables~\ref{tab-SMIN} and~\ref{tab-SMpred},
respectively.  In Figure~\ref{fig-chiex} the observed value of
$\Delta\chi^2 \equiv \chi^2 - \chi^2_{\mathrm{min}}$ as a function of
$\MH$ is plotted for this fit including all high-$Q^2$ results.  The
solid curve is the result using ZFITTER, and corresponds to the last
column of Table~\ref{tab-BIGFIT}.  The shaded band represents the
uncertainty due to uncalculated higher-order corrections, as estimated
by ZFITTER.

The 95\% confidence level upper limit on $\MH$ (taking the band into
account) is $144~\GeV$.  The 95\% C.L. lower limit on $\MH$ of
114.4~\GeV{} obtained from direct searches\cite{LEPSMHIGGS} is not
used in the determination of this limit.  Including it increases the
limit to $182~\GeV$.  Also shown is the result (dashed curve) obtained
when using $\Delta\alpha^{(5)}_{\mathrm{had}}(\MZ^2)$ of
Reference~\citen{bib-Troconiz-Yndurain-2004}.

\begin{table}[htbp]
\renewcommand{\arraystretch}{1.5}
  \begin{center}
\begin{tabular}{|c||c|c|c|c|c|}
\hline
&     - 1 -              &      - 2 -             &    - 3 -               &     - 4 -             \\
& all Z-pole             & all Z-pole data        & all Z-pole data        & all Z-pole data       \\[-3mm]
& data                   &    plus   $\Mt$        & plus $\MW$, $\GW$      & plus $\Mt,\MW,\GW$    \\
\hline
\hline
$\Mt$\hfill[\GeV] 
& $173^{+13 }_{-10}$     & $170.9^{+1.8}_{-1.8}$  & $179^{+12}_{- 9}$      & $171.3^{+1.7}_{-1.7}$  \\
$\MH$\hfill[\GeV] 
& $111^{+190}_{-60}$     & $ 99^{+52}_{-35}$      & $145^{+240}_{-81}$     & $ 76^{+33}_{-24}$      \\
$\log_{10}(\MH/\GeV)$  
& $2.05^{+0.43}_{-0.34}$ & $2.00^{+0.18}_{-0.19}$ & $2.16^{+0.42}_{-0.35}$ & $1.88^{+0.16}_{-0.17}$ \\
$\alfmz$          
& $0.1190\pm 0.0027$     & $0.1189\pm0.0027$      & $0.1190\pm 0.0028$     & $0.1185\pm 0.0026$     \\
\hline
$\chi^2$/d.o.f.{} ($P$)
& $16.0/10~(9.9\%)$      & $16.0/11~(14\%)$       & $17.4/12~(14\%)$       & $18.2/13~(15\%)$       \\
\hline
\hline
$\swsqeffl$
& $\pz0.23149$           & $\pz0.23149$           & $\pz0.23143$           & $\pz0.23138$ \\[-1mm]
& $\pm0.00016$           & $\pm0.00016$           & $\pm0.00014$           & $\pm0.00013$ \\
$\swsq$     
& $\pz0.22331$           & $\pz0.22338$           & $\pz0.22289$           & $\pz0.22311$ \\[-1mm]
& $\pm0.00062$           & $\pm0.00038$           & $\pm0.00038$           & $\pm0.00029$ \\
$\MW$\hfill[\GeV]
& $80.363\pm0.032$       & $80.360\pm0.020$       & $80.385\pm0.020$       & $80.374\pm0.015$   \\
\hline
\end{tabular}
\end{center}
\caption[]{ Results of the fits to: (1) all Z-pole data ({\LEPI} and
  SLD), (2) all Z-pole data plus direct $\Mt$ determination, (3) all
  Z-pole data plus direct $\MW$ and $\GW$ determinations, (4) all
  Z-pole data plus direct $\Mt,\MW,\GW$ determinations (i.e., all
  high-$Q^2$ results).  As the sensitivity to $\MH$ is logarithmic,
  both $\MH$ as well as $\log_{10}(\MH/\GeV)$ are quoted.  The bottom part
  of the table lists derived results for $\swsqeffl$, $\swsq$ and
  $\MW$.  See text for a discussion of theoretical errors not included
  in the errors above.  }
\label{tab-BIGFIT}
\renewcommand{\arraystretch}{1.0}
\end{table}

\begin{table}[htbp]
\begin{center}
  \renewcommand{\arraystretch}{1.30}
\begin{tabular}{|ll||r||r|r|l|}
\hline
 && {Measurement with}  & {Standard Model} & {Pull}  \\
 && {Total Error}       & {High-$Q^2$ Fit} & {    }  \\
\hline
\hline
&APV~\cite{QWCs:theo:2003:new}
                &                        &                      &       \\
\hline
&$\QWCs$        & $-72.74\pm0.46\pzz\pz$ & $-72.899\pm0.032\pzz$& $0.3$ \\
\hline
\hline
&M{\o}ller~\cite{E158RunI, *E158RunI+II+III}
                &                        &                      &        \\
\hline
&$\swsqMSb$     & $0.2330\pm0.0015\pz$   & $0.23109\pm0.00013$  & $1.3$  \\
\hline
\hline
&$\nu$N~\cite{bib-NuTeV-final}
                &                        &                      &        \\
\hline
&$\gnlq^2$      & $0.30005\pm0.00137$    & $0.30391\pm0.00016$  & $2.8$  \\
&$\gnrq^2$      & $0.03076\pm0.00110$    & $0.03011\pm0.00003$  & $0.6$  \\
\hline
\end{tabular}\end{center}
\caption[]{ Summary of measurements performed in low-$Q^2$ reactions,
  namely atomic parity violation, $e^-e^-$ M{\o}ller scattering and
  neutrino-nucleon scattering.  The SM results and the pulls
  (difference between measurement and fit in units of the total
  measurement error) are derived from the SM fit including all
  high-$Q^2$ data (Table~\ref{tab-BIGFIT}, column~4) with the Higgs
  mass treated as a free parameter.}
\label{tab-SMpred}
\end{table}

\begin{figure}[p]
\vspace*{-2.0cm}
\begin{center}
\includegraphics[height=21cm]{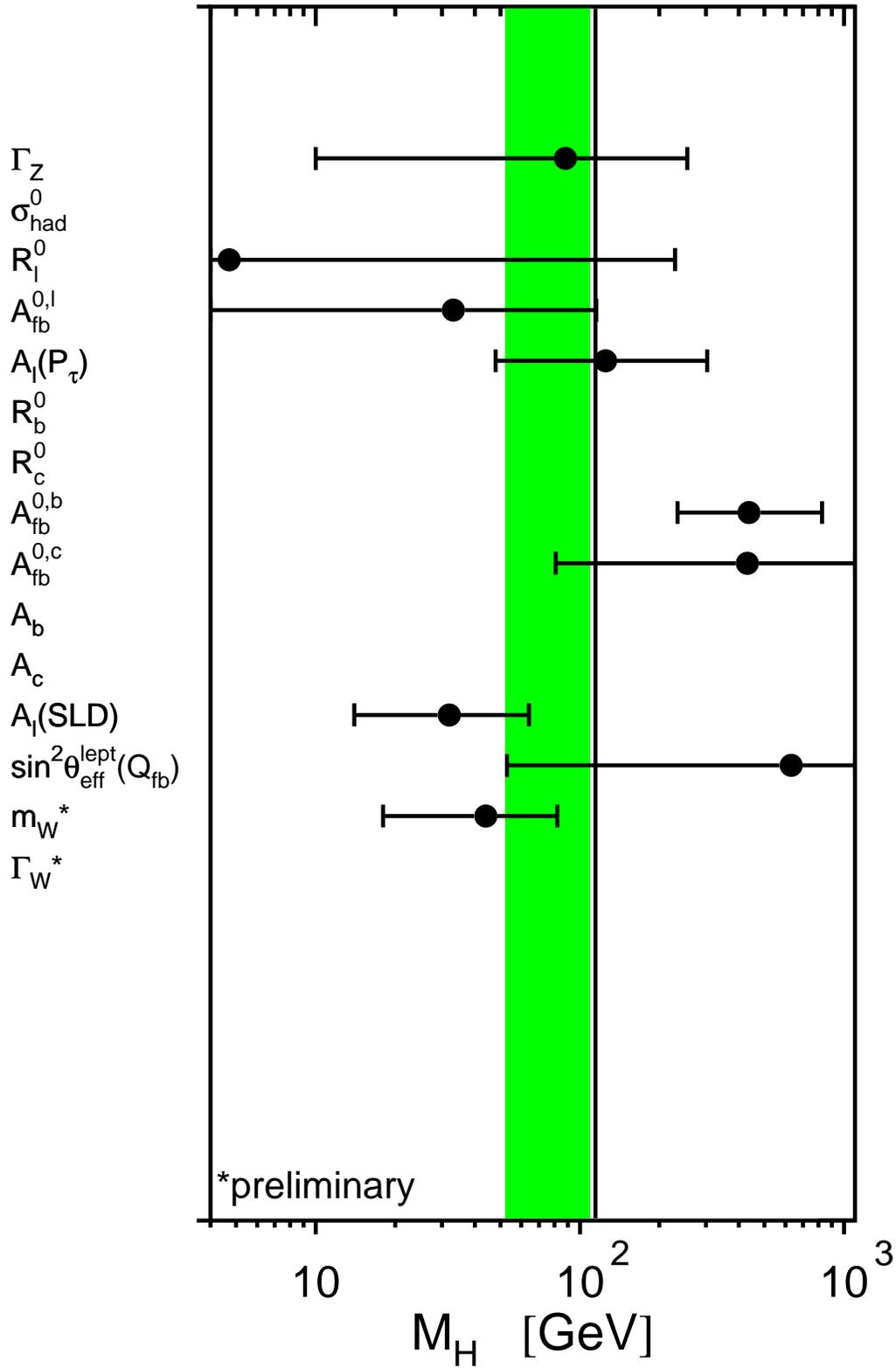}
\end{center}
\vspace*{-0.6cm}
\caption[]{ Constraints on the mass of the Higgs boson from each
  pseudo-observable. The Higgs-boson mass and its 68\% CL uncertainty
  is obtained from a five-parameter SM fit to the observable,
  constraining
  $\Delta\alpha^{(5)}_{\mathrm{had}}(\MZ^2)=0.02758\pm0.00035$,
  $\alfmz=0.118\pm0.003$, $\MZ=91.1875\pm0.0021~\GeV$ and
  $\Mt=170.9\pm1.8~\GeV$.  Because of these four common constraints
  the resulting Higgs-boson mass values are highly correlated.  The
  shaded band denotes the overall constraint on the mass of the Higgs
  boson derived from all pseudo-observables including the above four
  SM parameters as reported in the last column of
  Table~\ref{tab-BIGFIT}. The vertical line denotes the 95\% CL lower
  limit from the direct search for the Higgs boson. Results are only
  shown for constraints falling in the range of values shown. }
\label{fig-higgs-obs}
\end{figure}

Given the constraints on the other four SM input parameters, each
observable is equivalent to a constraint on the mass of the SM Higgs
boson. The constraints on the mass of the SM Higgs boson resulting
from each observable are compared in Figure~\ref{fig-higgs-obs}.  For
very low Higgs-masses, these constraints are qualitative only as the
effects of real Higgs-strahlung, neither included in the experimental
analyses nor in the SM calculations of expectations, may then become
sizeable~\cite{Kawamoto:2004pi}.
Besides the measurement of the W mass, the most sensitive measurements
are the asymmetries, \ie, $\swsqeffl$.  A reduced uncertainty for the
value of $\alpha(\MZ^2)$ would therefore result in an improved
constraint on $\log\MH$ and thus $\MH$, as already shown in
Figures~\ref{fig-gllsef} and \ref{fig-chiex}.


\clearpage

\boldmath
\section{Conclusions}
\label{sec-Conc}
\unboldmath

The preliminary and published results from the LEP, SLD and Tevatron
experiments, and their combinations, test the Standard Model (SM)
successfully at the highest interaction energies.  The combination of
the many precise electroweak results
yields stringent constraints on the SM and its free parameters.  Most
measurements agree well with the predictions.  The spread in values of
the various determinations of the effective electroweak mixing angle
in asymmetry measurements at the Z pole is somewhat larger than
expected~\cite{bib-Z-pole}.  
\boldmath
\section*{Prospects for the Future}
\unboldmath

The measurements from data taken at or near the Z resonance, both at
LEP as well as at SLC, are final and published~\cite{bib-Z-pole}.
Improvements in accuracy will therefore take place in the high energy
data (\LEPII), where each experiment has accumulated about
700~pb$^{-1}$ of data, and of course at the Tevatron.  The
measurements of $\MW$ are likely to reach a precision not too far from
the uncertainty on the prediction obtained via the radiative
corrections of the Z-pole data, providing an important test of the
Standard Model.
\section*{Acknowledgements}

We would like to thank the CERN accelerator divisions for the
efficient operation of the LEP accelerator, the precise information on
the absolute energy scale and their close cooperation with the four
experiments.  We would also like to thank members of the SLD, CDF,
D\O, E-158 and NuTeV collaborations for useful discussions concerning
their results.  Finally, the results of the section on Standard Model
constraints would not be possible without the close collaboration of
many theorists.

\clearpage

\bibliographystyle{PhysRep}
\bibliography{s07_ew,common,gg,ff,smat,fsi,be,4f_s06,gc,mw}

\end{document}